\documentclass[a4paper,fleqn,usenatbib]{mnras}

\usepackage[T1]{fontenc}
\usepackage{ae,aecompl}

\usepackage{graphicx}	
\usepackage{amsmath}	
\usepackage{amssymb}	
\usepackage{subfigure}
\usepackage{multirow}
\usepackage{longtable}

\newcommand{\pcc}{\,{\rm cm}^{-3}}
\newcommand{\gcc}{\,{\rm g \, cm}^{-3}}

\newcommand{\kel}{\, {\rm K}}

\newcommand{\msun}{\, {\rm M}_\odot}
\newcommand{\zsun}{\, {\rm Z}_\odot}
\newcommand{\nh}{n_{\rm H}}

\newcommand{\mh}{m_{\rm H}}

\newcommand{\yr}{\, {\rm yr}}
\newcommand{\kms}{\, {\rm km \, s^{-1}}}

\newcommand{\myr}{\, {\rm Myr}}
\newcommand{\gyr}{\, {\rm Gyr}}

\newcommand{\fdest}{f_{\rm dest}}

\newcommand{\mstar}{M_*}
\newcommand{\mgas}{M_{\rm gas}}
\newcommand{\mdust}{M_{\rm dust}}
\newcommand{\mz}{M_Z}
\newcommand{\fret}{f_{\rm ret}}
\newcommand{\yz}{y_Z}
\newcommand{\ydust}{y_{\rm dust}}
\newcommand{\zgas}{Z_{\rm gas}}
\newcommand{\zdust}{Z_{\rm dust}}
\newcommand{\ztot}{Z_{\rm tot}}
\newcommand{\facc}{f_{\rm acc}}

\newcommand{\mclr}{M_{\rm clear}}

\title[Dust-to-metals ratio]{The impact of metallicity-dependent dust destruction on the dust-to-metals ratio in galaxies}

\author[Priestley et al.]{
  F. D. Priestley$^{1}$\thanks{Email: priestleyf@cardiff.ac.uk},
  I. De Looze$^{2,3}$ and
M. J. Barlow$^{3}$
\\
$^{1}$School of Physics and Astronomy, Cardiff University, Queen's Buildings, The Parade, Cardiff CF24 3AA, UK \\
$^{2}$Sterrenkundig Observatorium, Ghent University, Krijgslaan 281 - S9, 9000 Gent, Belgium\\
$^{3}$Department of Physics and Astronomy, University College London, Gower Street, London WC1E 6BT, UK\\
}

\date{Accepted XXX. Received YYY; in original form ZZZ}

\pubyear{2021}

\begin{document}
\label{firstpage}
\pagerange{\pageref{firstpage}--\pageref{lastpage}}
\maketitle

\begin{abstract}

  The ratio of the mass of interstellar dust to the total mass of metals (the dust-to-metals/DTM ratio) {tends to increase with metallicity}. This can be explained by the increasing efficiency of grain growth in the interstellar medium (ISM) at higher metallicities, with a corollary being that the low DTM ratios seen at low metallicities are due to inefficient stellar dust production. This interpretation assumes that the efficiency of dust destruction in the ISM is constant, whereas it might be expected to increase at low metallicity; the decreased cooling efficiency of low-metallicity gas should result in more {post-shock} dust destruction via thermal sputtering. We show that incorporating {a sufficiently strong} metallicity dependence into models of galaxy evolution removes the need for low stellar dust yields. The contribution of stellar sources to the overall dust budget may be significantly underestimated, and that of grain growth overestimated, {by models assuming a constant destruction efficiency.}

\end{abstract}




\section{Introduction}

The variation of the dust-to-metals (DTM) ratio with the total metal content of galaxies is often seen as strong evidence for grain growth in the interstellar medium (ISM) as the main driver of dust evolution. {Models often struggle to reproduce the observed distribution of galaxies in the DTM ratio-metallicity plane with efficient dust production {by core-collapse supernovae (CCSNe)}, leaving ISM growth as the only way to account for the observed dust masses \citep{devis2017,devis2019,galliano2021} {unless dust yields from lower-mass stars are substantially higher than is typically assumed}.} However, the required {limits on CCSN} dust yields {($\lesssim 0.1 \msun$ per CCSN)} are well below those observed in supernova remnants \citep{matsuura2015,delooze2017,delooze2019,chawner2019}, even after accounting for possible future destruction via reverse shocks \citep{delooze2017,priestley2019,niculescu2021}.

While models of ISM enrichment and evolution typically account for the metallicity-dependence of grain growth, in that the rate of accretion increases with the availability of gas-phase metals, the efficiency of dust destruction in the ISM is often assumed to be constant. This is potentially inaccurate; metal-poor gas cools less efficiently, thus remaining at high temperatures for longer, and increasing the amount of dust destroyed by thermal sputtering in shocked gas (\citealt{yamasawa2011}, hereafter Y11). Some studies of dust evolution \citep[e.g.][]{triani2020} do use a metallicity-dependent destruction efficiency, but to our knowledge, the impact of {such an evolving efficiency has not been investigated in detail. In this paper, we show that an increasing destruction efficiency at low metallicity {may remove} the tension between high stellar dust yields and the low DTM ratios seen in some galaxies, {if the dependence on metallicity is strong enough}.} Without accounting for this effect, the importance of stellar sources to the global dust budget {may be} significantly underestimated.

\section{Method}

We consider a {closed-box} model of galaxy evolution {with a single-phase ISM}, tracking four main quantities; the masses of gas ($\mgas$), stars ($\mstar$), gas-phase metals ($\mz$), and dust ($\mdust$). Initially-pristine gas is converted into stars, which return a fraction of that gas to the ISM, enriched with metals and dust under the instantaneous-recycling approximation. Gas-phase metals are accreted onto dust grains, and metals locked up in dust grains are released into the gas via dust destruction. The evolution of the system is given by
\begin{eqnarray}
  \frac{d \mgas}{dt} = - (1 - \fret) \, \Sigma \\
  \frac{d \mstar}{dt} = (1 - \fret) \, \Sigma \\
  \frac{d \mz}{dt} = (\fret \yz - \zgas) \Sigma + D - G \\
  \frac{d \mdust}{dt} = (\fret \ydust - \zdust) \Sigma - D + G
\end{eqnarray}
where $\Sigma$ is the star formation rate (SFR), $\fret$ is the fraction of mass returned by stars into the ISM, $\yz$ and $\ydust$ are the stellar yields of gas-phase metals and dust respectively (the fraction of mass returned in the form of either), $\zgas = \mz/\mgas$ and $\zdust = \mdust/\mgas$ are the gas- and dust-phase metallicities, and $D$ and $G$ are the dust destruction and growth rates respectively. The total metallicity of the `galaxy' is $\ztot = (\mz + \mdust)/\mgas = \zgas + \zdust$, {and the DTM ratio is $\mdust/(\mz + \mdust)$}. We assume initial values of $\mgas = 10^{10} \msun$ and $\mstar = \mz = \mdust = 0$, and evolve the model for $10 \gyr$, with an adaptive timestep of $0.01 \times {\rm min} \left(\mgas/\Sigma,\mz/G,\mdust/D\right)$. {As we are mainly interested in ratios of masses, rather than their absolute values, the initial value of $\mgas$ has no impact on our results.} {We take the \citet{asplund2009} value of $\zsun = 0.0134$ for solar metallicity.}

\begin{table}
  \centering
  \caption{Values of model parameters which are held constant.}
  \begin{tabular}{ccc}
    \hline
    Parameter & Value & Unit \\
    \hline
    $\fret$ & $0.1$ & - \\
    $\yz + \ydust$ & $0.1$ & - \\
    $A_g$ & $4.24 \times 10^{5}$ & cm$^2$ g$^{-1}$ \\
    $T_{\rm gas}$ & 20 & K \\
    $\mu_Z$ & 12 & $\mh$ \\
    $\mu$ & $2.33$ & $\mh$ \\
    $\nh$ & 100 & $\pcc$ \\
    $f_{\rm SN}$ & $0.01$ & $\msun^{-1}$ \\
    \hline
  \end{tabular}
  \label{tab:param}
\end{table}

\begin{figure*}
  \centering
  \subfigure{\includegraphics[width=\columnwidth]{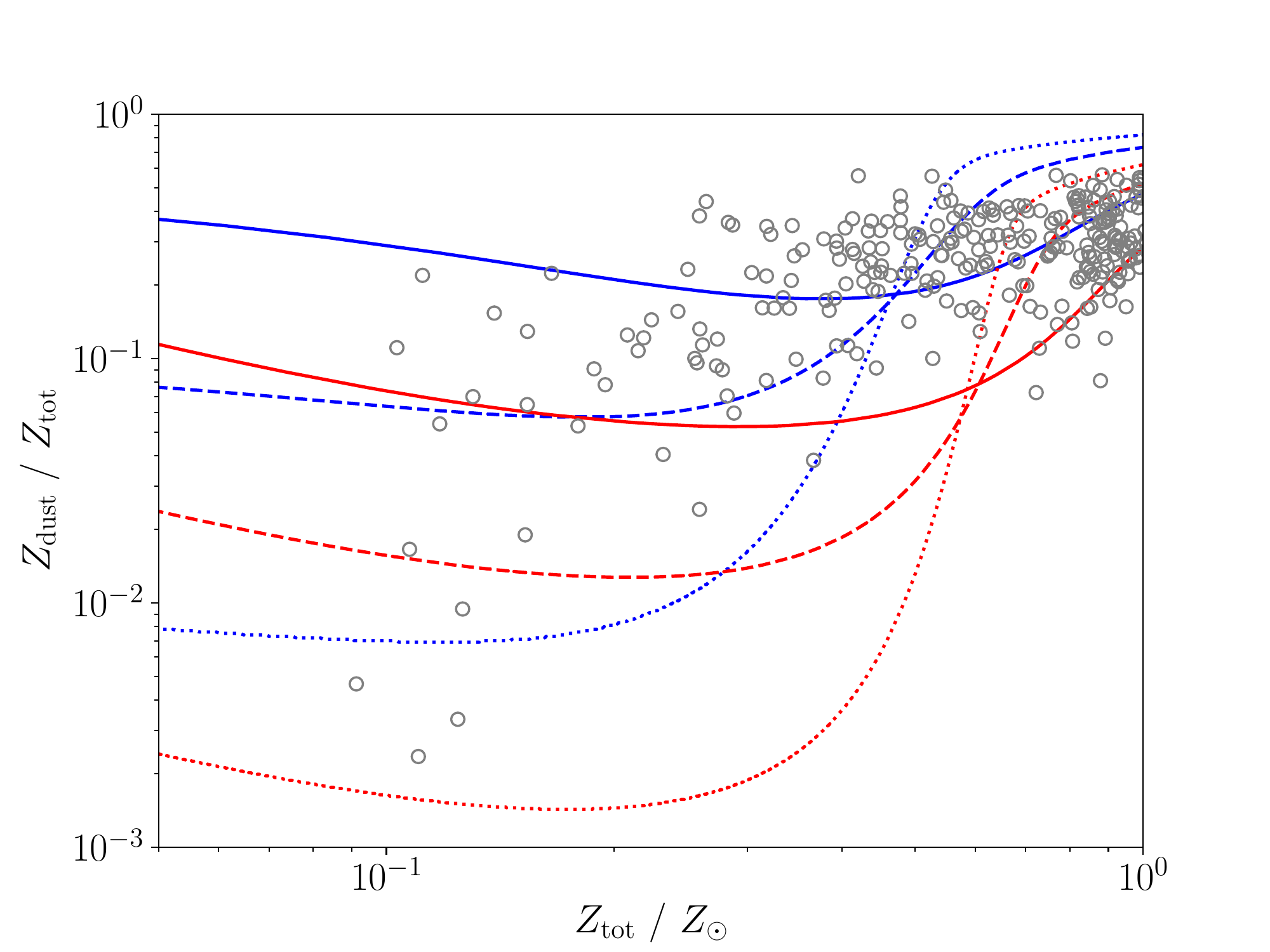}}\quad
  \subfigure{\includegraphics[width=\columnwidth]{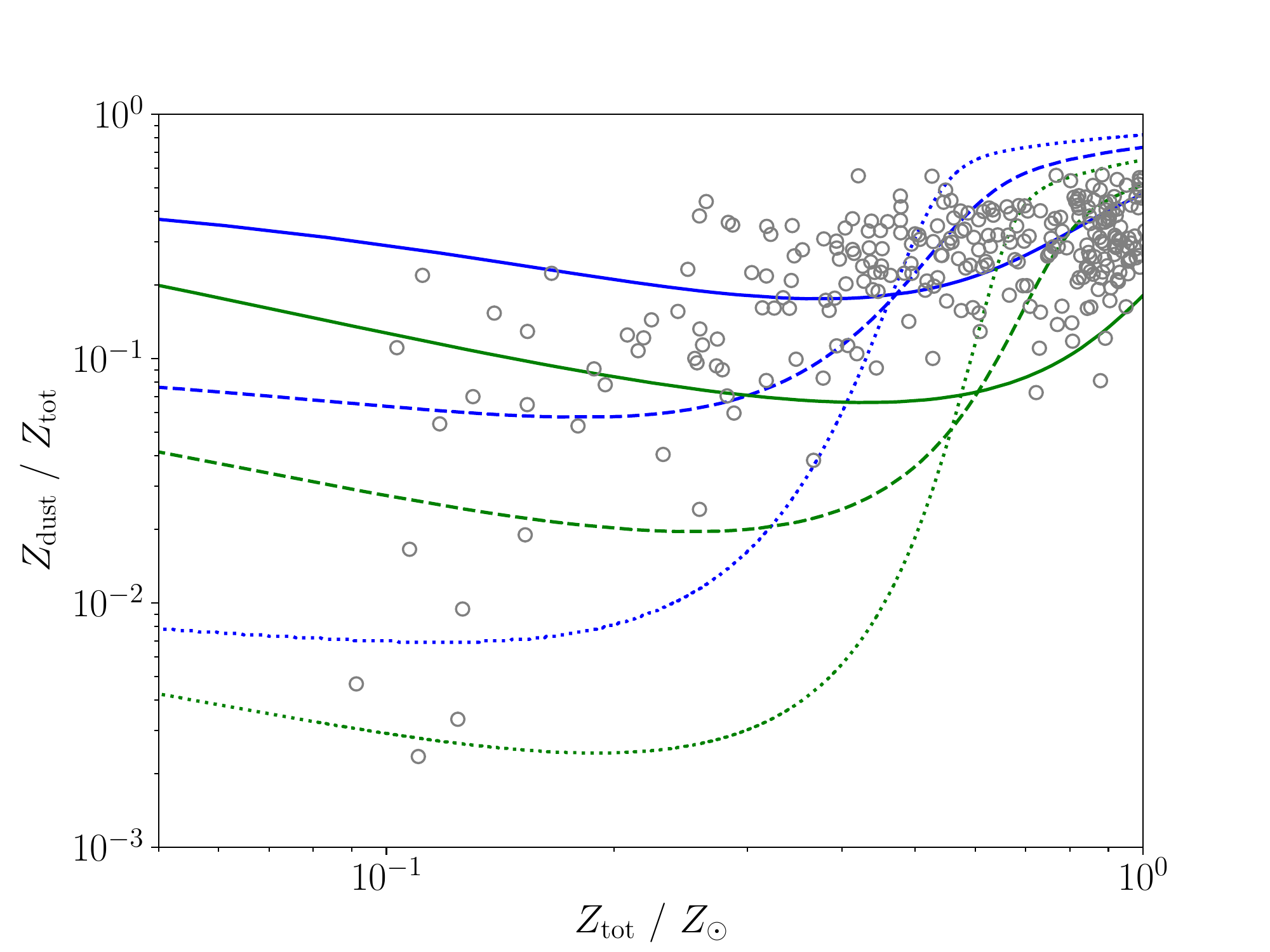}}\quad
  \caption{Evolution of the DTM ratio versus metallicity {for models with a $\ydust$ ($\facc$) of $0.05$ ($0.005$) (solid lines), $0.01$ ($0.01$) (dashed lines) and $0.001$ ($0.015$) (dotted lines). The left panel shows models with a constant $\mclr = 1000 \msun$ (blue lines) and with $\mclr$ given by Equation \ref{eq:mclr} (red lines). The right panel shows models with a constant $\mclr = 1000 \msun$ (blue lines) and with the Y11 $\mclr$ (green lines).} Observational data from \citet{galliano2021} are shown as grey open circles.}
  \label{fig:mclr}
\end{figure*}

\subsection{Star formation}

The SFR, $\Sigma$, is not in general a function of the other model parameters, as galaxies grow and evolve due to the accretion and ejection of gas, and occasional mergers \citep{triani2020,nanni2020}. We assume
\begin{equation}
  \Sigma = 2 \times 10^6 \left( \frac{\mgas}{10^{10} \msun} \right) \msun \myr^{-1},
  \label{eq:sfr}
\end{equation}
{which gives $1 \msun \yr^{-1}$ for a gas mass of $5 \times 10^9 \msun$, in line with Milky Way values \citep{dame1993,robitaille2010}, and produces an exponentially-declining SFR.} The galaxy scaling relations we are interested in are typically more sensitive to the assumptions made regarding the dust physics, rather than the star formation history \citep{devis2017,delooze2020}.

\subsection{Grain growth}

The increase in dust mass due to growth in the ISM is given by
\begin{equation}
  G = \facc \times A_g \mgas \zdust \times \sqrt{\frac{k_b T_{\rm gas}}{\mu_Z}} \mu \nh \zgas.
  \label{eq:grow}
\end{equation}
Here, $A_g$ is the grain surface area per unit mass, $k_b$ is the Boltzmann constant, $T_{\rm gas}$, $\nh$ and $\mu$ the gas temperature, number density and mean molecular mass in the phase of the ISM where grain growth occurs, and $\mu_Z$ is the mass of a typical accreting atom. $\facc$ is a parameter representing, among other things, the fraction of the total gas mass in which grain growth occurs, {and the sticking efficiency of collisions between atoms and grains}. The second term in Equation \ref{eq:grow} gives the total available grain surface area for accretion, and the third term is the {thermal velocity of accreting atoms multiplied by their mass density, giving the `mass flux' per unit area.}

\subsection{Dust destruction}

Along with astration and outflows, destruction in shocks driven by SNe is one of the main processes removing dust from the ISM. As individual SNe cannot be resolved even in hydrodynamical simulations of galaxy evolution, their effects are generally abstracted to a dust destruction timescale representing the typical lifetime of grains in the ISM \citep[e.g.][]{morgan2003}. We assume the rate of dust destruction is given by
\begin{equation}
  D = f_{\rm SN} \Sigma \times \zdust \mclr,
  \label{eq:dest}
\end{equation}
where $f_{\rm SN}$ is the number of SNe per unit mass of stars formed, $\Sigma$ is the SFR given in Equation \ref{eq:sfr}, and $\mclr$ is the gas mass `cleared' of dust per SN.

\subsection{Parameter choices}

The free parameters controlling grain growth and destruction in Equations \ref{eq:grow} and \ref{eq:dest} are degenerate, so we fix all but one in each case to physically-reasonable values. For grain growth, we assume that this process occurs in molecular gas with $\nh = 100 \pcc$, $T_{\rm gas} = 20 \kel$ and $\mu = 2.33 \mh$, that the accreting atoms have $\mu_Z = 12 \mh$, and that the area-to-mass ratio of dust grains is that of a \citet{mathis1977} distribution with a bulk density of $2 \gcc$ ($4.24 \times 10^{5} \, {\rm cm^2 \, g^{-1}})$. For $f_{\rm SN}$, we assume a value of $0.01 \msun^{-1}$ consistent with a \citet{salpeter1955} initial mass function and a minimum SN mass of $\sim 10 \msun$. We additionally assume a returned mass fraction $\fret = 0.1$ and a total metal yield $\yz + \ydust = 0.1$, in line with stellar evolution models \citep{woosley1995}. {The metal yields of both asymptotic giant branch (AGB) stars \citep{cristallo2009} and CCSNe from $\lesssim 30 \msun$ progenitors\footnote{Higher-mass progenitors account for only $\sim 20\%$ of CCSNe for a \citet{salpeter1955} power law, so are relatively unimportant as a source of metals.} \citep{limongi2018} appear to vary only modestly with metallicity, so we take these values as constant.} Values of the fixed parameters are listed in Table \ref{tab:param}.

The remaining free parameters are the grain growth efficiency $\facc$, the mass of gas cleared per SN $\mclr$, and the distribution of newly-produced metals between the gas and dust phases, $\yz$ and $\ydust$. We assume that $\facc$ and the balance between $\yz$ and $\ydust$ both remain constant with metallicity. The accretion efficiency is not necessarily constant, {as both the fraction of the accreting phase of the ISM \citep{mattsson2012} and the ability of grains to accrete atoms \citep{ferrara2016} may change over time, but our results are not greatly affected by introducing additional dependencies (e.g. $\facc \propto \Sigma$).} {As with the total metal yields, theoretical DTM ratios of both AGB outflows \citep{nanni2013} and CCSN ejecta from lower-mass progenitors \citep{marassi2019} do not appear to vary significantly with metallicity.}

The cleared gas mass per SN, $\mclr$, is typically estimated to be $\sim 1000 \msun$ under local ISM conditions \citep{jones1994,hu2019}, although there are arguments in favour of a lower value \citep{martinez2019,priestley2021,ferrara2021}. $\mclr$ is also likely to vary with metallicity; at lower $\zgas$, gas cools less efficiently, and thus remains at the high temperatures necessary for efficient thermal sputtering for longer. {Based on the {variation with metallicity of the dust destruction efficiency} {derived in} Appendix \ref{sec:scaling}, we consider}
\begin{equation}
  \mclr = \frac{8143 \msun}{1 + \zgas / 0.14 \zsun}
  \label{eq:mclr}
\end{equation}
{{which gives $\mclr \sim 1000 \msun$ at $\zgas = \zsun$. We also investigate models with a constant $\mclr = 1000 \msun$ for comparison.}

{Equation \ref{eq:mclr} may overestimate the strength of the dependence of $\mclr$ on $\zgas$} - Y11, using {a more detailed model of dust destruction}, find {a much weaker scaling ($\mclr \sim Z^{-0.298}$ at high $Z$)}. We use their eq. 8 for $\mclr$, with $n = 1 \pcc$, to investigate how the implementation of the metallicity dependence affects our results.}

\section{Results}

{Figure \ref{fig:mclr} shows the evolution of the DTM ratio for models with a constant $\mclr$, {with $\ydust$ varied between the observationally-favoured $0.05$, and the $0.001$ determined from previous work on the DTM ratio \citep[e.g.][]{galliano2021}\footnote{The dust yield per CCSN, including any AGB-formed dust, is $\ydust \times \fret / f_{\rm SN} = 10 \ydust \msun$, so these values correspond to $0.5$ and $0.01 \msun$ per CCSN respectively.}. $\facc$ is chosen to reproduce the observed rise in DTM ratio between $0.1$ and $0.5 \zsun$, and the value of $\sim 0.5$ at $\ztot = \zsun$.} At low values of $\ztot$, the efficiency of both destruction and growth in the ISM is low, and the DTM ratio tends towards the intrinsic value from stellar production. Reproducing the low values seen at $\ztot \sim 0.1 \zsun$ then requires $\ydust \lesssim 0.01$, which corresponds to a dust yield per CCSN of $\lesssim 0.1 \msun$, as found by \citet{galliano2021} and other previous studies.}

{The {left} panel of Figure \ref{fig:mclr} shows the impact of a metallicity-dependent $\mclr$, {implemented according to Equation \ref{eq:mclr}.} The increased dust destruction at low metallicity counteracts the high stellar dust yields, {so that the DTM ratio at $0.1 \zsun$ is significantly lower than the intrinsic stellar value. An observed value of $\zdust/\ztot \sim 0.01$ is then consistent with an average CCSN yield of $0.1 \msun$, and the low-$Z$ data as a whole do not rule out the $\sim 0.5 \msun$ dust masses seen in supernova remnants.} {Using instead the Y11 prescription for $\mclr$, shown in the right panel of Figure \ref{fig:mclr}, the weaker metallicity dependence results in a more modest decrease in the low-$Z$ DTM ratio compared to the constant-$\mclr$ models, but the observed DTM ratios of $\lesssim 0.1$ can still be reproduced by models with substantial ($\gtrsim 0.1 \msun$ per CCSN) stellar dust yields.}}

\section{Discussion \& Conclusions}
\label{sec:discussion}

{It is often claimed that the low DTM ratios seen in some galaxies require similarly low dust yields {from CCSNe}. We have shown that this {interpretation is affected by} the assumption of a constant dust destruction efficiency. If preexisting dust grains are more easily destroyed in lower-metallicity gas, {and the resulting increase in destruction efficiency depends strongly enough on metallicity}, there is no tension between the large dust masses found in supernova remnants and the relatively small dust masses in low-metallicity galaxies.}

{As we note in Appendix \ref{sec:scaling}, Equation \ref{eq:mclr} is likely to overestimate the strength of the metallicity dependence of $\mclr$ by neglecting kinetic sputtering. This explains the weaker dependence found by Y11, who include this process. However, Y11 themselves neglect grain shattering, which can have an enormous impact on dust destruction rates \citep{kirchschlager2019}. While shattering should not be greatly affected by the gas-phase metallicity, we would expect it to redistribute mass from large to small grains, and therefore increase the importance of thermal compared to kinetic sputtering\footnote{Large grains, due to their high inertia, are more strongly affected by kinetic sputtering, whereas small grains are typically well-coupled to the gas velocity but rapidly eroded by hot gas due to their higher surface area to volume ratios.}. A more complete treatment of dust destruction as a function of metallicity may therefore be closer to the scaling in Equation \ref{eq:mclr} than to the Y11 results.}

{A limitation of our model (and most other single-phase models) is the assumption that newly-produced metals and dust are instantaneously mixed with the entire gas reservoir. As dust production is likely to occur in the vicinity of star-forming regions, either due to stellar dust production or growth in dense molecular gas, the ISM swept up by CCSNe is likely to be more dust-enriched than average. Equation \ref{eq:dest} will then underestimate the actual rate of dust destruction. This effect is particularly severe at low metallicity, where the instantaneous-mixing assumption results in such a low $\zdust$ that destruction by SNe is negligible, and DTM ratios {\it increase} from $\ztot \sim 0.01 \zsun$ to $\ztot = 0$ (Figure \ref{fig:mclr}). As such, we may still be underestimating dust destruction at low metallicity. On the other hand, low-metallicity dwarf galaxies are only found to be chemically inhomogeneous by factors of a few \citep{lebouteiller2009,lebouteiller2013,james2016,james2020}, so mixing may be efficient enough to limit the importance of this effect.}

A metallicity-dependent dust destruction efficiency is far from the only possible cause of an increase in the DTM ratio. While we have assumed stellar dust yields are constant, as indicated by models, the current theoretical understanding of dust formation appears to be incomplete \citep{wesson2015,bevan2016,priestley2020b} so this may well be inaccurate. \citet{devis2021} have recently reconciled high stellar dust yields with low DTM ratios by invoking photofragmentation, although as this process only affects carbon grains, it is unclear whether it can achieve the required reduction in dust mass. {As our proposed mechanism is based on comparatively well-understood physical processes, which must be in operation to some extent in the ISM, we regard it as a more promising explanation of the observational data.}

{By allowing high stellar dust yields without an increase in the destruction rate, at least at higher metallicities, an evolving $\mclr$ inevitably results in a greater importance for stellar dust production compared to ISM growth. {With $\mclr$ given by Equation \ref{eq:mclr}, the lowest observed DTM ratios can be reproduced with a value of $\ydust$ an order of magnitude larger than that required by a constant $\mclr$ model. This results in a corresponding order of magnitude increase in the contribution of stellar sources to the overall dust budget.} Depending on the values of the other parameters in our model, this increase may not be of huge importance - the majority of the dust mass after $10 \gyr$ is still ISM-grown for {our models, because the adopted value (or normalisation) of $\mclr$ results in CCSNe destroying more dust than is formed in stars by the time $\ztot \sim \zsun$}. However, {the common assertion that stellar dust production is unimportant compared to grain growth, in both low- and high-redshift galaxies, may be reliant on the assumption of a constant destruction efficiency. We suggest that models making this assumption are reevaluated in order to determine its impact.}}

\section*{Acknowledgements}

FDP is funded by the Science and Technology Facilities Council. IDL acknowledges support from European Research Council (ERC) starting grant 851622 DustOrigin. MJB acknowledges support from the ERC grant SNDUST ERC-2015-AdG-694520.

\section*{Data Availability}

The data underlying this article will be made available upon request. A Fortran implementation of the model is available at \url{www.github.com/fpriestley/galaxy}.




\bibliographystyle{mnras}
\bibliography{dustmetal}

\begin{thebibliography}{}
\makeatletter
\relax
\def\mn@urlcharsother{\let\do\@makeother \do\$\do\&\do\#\do\^\do\_\do\%\do\~}
\def\mn@doi{\begingroup\mn@urlcharsother \@ifnextchar [ {\mn@doi@}
  {\mn@doi@[]}}
\def\mn@doi@[#1]#2{\def\@tempa{#1}\ifx\@tempa\@empty \href
  {http://dx.doi.org/#2} {doi:#2}\else \href {http://dx.doi.org/#2} {#1}\fi
  \endgroup}
\def\mn@eprint#1#2{\mn@eprint@#1:#2::\@nil}
\def\mn@eprint@arXiv#1{\href {http://arxiv.org/abs/#1} {{\tt arXiv:#1}}}
\def\mn@eprint@dblp#1{\href {http://dblp.uni-trier.de/rec/bibtex/#1.xml}
  {dblp:#1}}
\def\mn@eprint@#1:#2:#3:#4\@nil{\def\@tempa {#1}\def\@tempb {#2}\def\@tempc
  {#3}\ifx \@tempc \@empty \let \@tempc \@tempb \let \@tempb \@tempa \fi \ifx
  \@tempb \@empty \def\@tempb {arXiv}\fi \@ifundefined
  {mn@eprint@\@tempb}{\@tempb:\@tempc}{\expandafter \expandafter \csname
  mn@eprint@\@tempb\endcsname \expandafter{\@tempc}}}

\bibitem[\protect\citeauthoryear{{Asplund}, {Grevesse}, {Sauval}  \&
  {Scott}}{{Asplund} et~al.}{2009}]{asplund2009}
{Asplund} M.,  {Grevesse} N.,  {Sauval} A.~J.,   {Scott} P.,  2009, \mn@doi
  [\araa] {10.1146/annurev.astro.46.060407.145222}, \href
  {http://adsabs.harvard.edu/abs/2009ARA%26A..47..481A} {47, 481}

\bibitem[\protect\citeauthoryear{{Bevan} \& {Barlow}}{{Bevan} \&
  {Barlow}}{2016}]{bevan2016}
{Bevan} A.,  {Barlow} M.~J.,  2016, \mn@doi [\mnras] {10.1093/mnras/stv2651},
  \href {http://adsabs.harvard.edu/abs/2016MNRAS.456.1269B} {456, 1269}

\bibitem[\protect\citeauthoryear{{Biscaro} \& {Cherchneff}}{{Biscaro} \&
  {Cherchneff}}{2016}]{biscaro2016}
{Biscaro} C.,  {Cherchneff} I.,  2016, \mn@doi [\aap]
  {10.1051/0004-6361/201527769}, \href
  {http://adsabs.harvard.edu/abs/2016A%26A...589A.132B} {589, A132}

\bibitem[\protect\citeauthoryear{{Chawner} et~al.,}{{Chawner}
  et~al.}{2019}]{chawner2019}
{Chawner} H.,  et~al., 2019, \mn@doi [\mnras] {10.1093/mnras/sty2942}, \href
  {http://adsabs.harvard.edu/abs/2019MNRAS.483...70C} {483, 70}

\bibitem[\protect\citeauthoryear{{Chawner} et~al.,}{{Chawner}
  et~al.}{2020}]{chawner2020b}
{Chawner} H.,  et~al., 2020, \mn@doi [\mnras] {10.1093/mnras/staa2925}, \href
  {https://ui.adsabs.harvard.edu/abs/2020MNRAS.499.5665C} {499, 5665}

\bibitem[\protect\citeauthoryear{{Cristallo}, {Straniero}, {Gallino},
  {Piersanti}, {Dom{\'\i}nguez}  \& {Lederer}}{{Cristallo}
  et~al.}{2009}]{cristallo2009}
{Cristallo} S.,  {Straniero} O.,  {Gallino} R.,  {Piersanti} L.,
  {Dom{\'\i}nguez} I.,   {Lederer} M.~T.,  2009, \mn@doi [\apj]
  {10.1088/0004-637X/696/1/797}, \href
  {https://ui.adsabs.harvard.edu/abs/2009ApJ...696..797C} {696, 797}

\bibitem[\protect\citeauthoryear{{Dame}}{{Dame}}{1993}]{dame1993}
{Dame} T.~M.,  1993, in {Holt} S.~S.,  {Verter} F.,  eds,  American Institute
  of Physics Conference Series Vol. 278, Back to the Galaxy. pp 267--278,
  \mn@doi{10.1063/1.43985}

\bibitem[\protect\citeauthoryear{{De Looze}, {Barlow}, {Swinyard}, {Rho},
  {Gomez}, {Matsuura}  \& {Wesson}}{{De Looze} et~al.}{2017}]{delooze2017}
{De Looze} I.,  {Barlow} M.~J.,  {Swinyard} B.~M.,  {Rho} J.,  {Gomez} H.~L.,
  {Matsuura} M.,   {Wesson} R.,  2017, \mn@doi [\mnras]
  {10.1093/mnras/stw2837}, \href
  {http://adsabs.harvard.edu/abs/2017MNRAS.465.3309D} {465, 3309}

\bibitem[\protect\citeauthoryear{{De Looze} et~al.,}{{De Looze}
  et~al.}{2019}]{delooze2019}
{De Looze} I.,  et~al., 2019, \mn@doi [\mnras] {10.1093/mnras/stz1533}, \href
  {https://ui.adsabs.harvard.edu/abs/2019MNRAS.488..164D} {488, 164}

\bibitem[\protect\citeauthoryear{{De Looze} et~al.,}{{De Looze}
  et~al.}{2020}]{delooze2020}
{De Looze} I.,  et~al., 2020, \mn@doi [\mnras] {10.1093/mnras/staa1496}, \href
  {https://ui.adsabs.harvard.edu/abs/2020MNRAS.496.3668D} {496, 3668}

\bibitem[\protect\citeauthoryear{{De Vis} et~al.,}{{De Vis}
  et~al.}{2017}]{devis2017}
{De Vis} P.,  et~al., 2017, \mn@doi [\mnras] {10.1093/mnras/stx981}, \href
  {https://ui.adsabs.harvard.edu/abs/2017MNRAS.471.1743D} {471, 1743}

\bibitem[\protect\citeauthoryear{{De Vis} et~al.,}{{De Vis}
  et~al.}{2019}]{devis2019}
{De Vis} P.,  et~al., 2019, \mn@doi [\aap] {10.1051/0004-6361/201834444}, \href
  {https://ui.adsabs.harvard.edu/abs/2019A&A...623A...5D} {623, A5}

\bibitem[\protect\citeauthoryear{{De Vis}, {Maddox}, {Gomez}, {Jones}  \&
  {Dunne}}{{De Vis} et~al.}{2021}]{devis2021}
{De Vis} P.,  {Maddox} S.~J.,  {Gomez} H.~L.,  {Jones} A.~P.,   {Dunne} L.,
  2021, \mn@doi [\mnras] {10.1093/mnras/stab1604}, \href
  {https://ui.adsabs.harvard.edu/abs/2021MNRAS.505.3228D} {505, 3228}

\bibitem[\protect\citeauthoryear{{Ferrara} \& {Peroux}}{{Ferrara} \&
  {Peroux}}{2021}]{ferrara2021}
{Ferrara} A.,  {Peroux} C.,  2021, \mn@doi [\mnras] {10.1093/mnras/stab761},
  \href {https://ui.adsabs.harvard.edu/abs/2021MNRAS.503.4537F} {503, 4537}

\bibitem[\protect\citeauthoryear{{Ferrara}, {Viti}  \& {Ceccarelli}}{{Ferrara}
  et~al.}{2016}]{ferrara2016}
{Ferrara} A.,  {Viti} S.,   {Ceccarelli} C.,  2016, \mn@doi [\mnras]
  {10.1093/mnrasl/slw165}, \href
  {https://ui.adsabs.harvard.edu/abs/2016MNRAS.463L.112F} {463, L112}

\bibitem[\protect\citeauthoryear{{Galliano} et~al.,}{{Galliano}
  et~al.}{2021}]{galliano2021}
{Galliano} F.,  et~al., 2021, \mn@doi [\aap] {10.1051/0004-6361/202039701},
  \href {https://ui.adsabs.harvard.edu/abs/2021A&A...649A..18G} {649, A18}

\bibitem[\protect\citeauthoryear{{Hu}, {Zhukovska}, {Somerville}  \&
  {Naab}}{{Hu} et~al.}{2019}]{hu2019}
{Hu} C.-Y.,  {Zhukovska} S.,  {Somerville} R.~S.,   {Naab} T.,  2019, \mn@doi
  [\mnras] {10.1093/mnras/stz1481}, \href
  {https://ui.adsabs.harvard.edu/abs/2019MNRAS.487.3252H} {487, 3252}

\bibitem[\protect\citeauthoryear{{James}, {Auger}, {Aloisi}, {Calzetti}  \&
  {Kewley}}{{James} et~al.}{2016}]{james2016}
{James} B.~L.,  {Auger} M.,  {Aloisi} A.,  {Calzetti} D.,   {Kewley} L.,  2016,
  \mn@doi [\apj] {10.3847/0004-637X/816/1/40}, \href
  {https://ui.adsabs.harvard.edu/abs/2016ApJ...816...40J} {816, 40}

\bibitem[\protect\citeauthoryear{{James}, {Kumari}, {Emerick}, {Koposov},
  {McQuinn}, {Stark}, {Belokurov}  \& {Maiolino}}{{James}
  et~al.}{2020}]{james2020}
{James} B.~L.,  {Kumari} N.,  {Emerick} A.,  {Koposov} S.~E.,  {McQuinn} K.
  B.~W.,  {Stark} D.~P.,  {Belokurov} V.,   {Maiolino} R.,  2020, \mn@doi
  [\mnras] {10.1093/mnras/staa1280}, \href
  {https://ui.adsabs.harvard.edu/abs/2020MNRAS.495.2564J} {495, 2564}

\bibitem[\protect\citeauthoryear{{Jones}, {Tielens}, {Hollenbach}  \&
  {McKee}}{{Jones} et~al.}{1994}]{jones1994}
{Jones} A.~P.,  {Tielens} A.~G.~G.~M.,  {Hollenbach} D.~J.,   {McKee} C.~F.,
  1994, \mn@doi [\apj] {10.1086/174689}, \href
  {https://ui.adsabs.harvard.edu/abs/1994ApJ...433..797J} {433, 797}

\bibitem[\protect\citeauthoryear{{Kirchschlager}, {Schmidt}, {Barlow},
  {Fogerty}, {Bevan}  \& {Priestley}}{{Kirchschlager}
  et~al.}{2019}]{kirchschlager2019}
{Kirchschlager} F.,  {Schmidt} F.~D.,  {Barlow} M.~J.,  {Fogerty} E.~L.,
  {Bevan} A.,   {Priestley} F.~D.,  2019, \mn@doi [\mnras]
  {10.1093/mnras/stz2399}, \href
  {https://ui.adsabs.harvard.edu/abs/2019MNRAS.489.4465K} {489, 4465}

\bibitem[\protect\citeauthoryear{{Lebouteiller}, {Kunth}, {Thuan}  \&
  {D{\'e}sert}}{{Lebouteiller} et~al.}{2009}]{lebouteiller2009}
{Lebouteiller} V.,  {Kunth} D.,  {Thuan} T.~X.,   {D{\'e}sert} J.~M.,  2009,
  \mn@doi [\aap] {10.1051/0004-6361:200811089}, \href
  {https://ui.adsabs.harvard.edu/abs/2009A&A...494..915L} {494, 915}

\bibitem[\protect\citeauthoryear{{Lebouteiller}, {Heap}, {Hubeny}  \&
  {Kunth}}{{Lebouteiller} et~al.}{2013}]{lebouteiller2013}
{Lebouteiller} V.,  {Heap} S.,  {Hubeny} I.,   {Kunth} D.,  2013, \mn@doi
  [\aap] {10.1051/0004-6361/201220948}, \href
  {https://ui.adsabs.harvard.edu/abs/2013A&A...553A..16L} {553, A16}

\bibitem[\protect\citeauthoryear{{Limongi} \& {Chieffi}}{{Limongi} \&
  {Chieffi}}{2018}]{limongi2018}
{Limongi} M.,  {Chieffi} A.,  2018, \mn@doi [\apjs] {10.3847/1538-4365/aacb24},
  \href {https://ui.adsabs.harvard.edu/abs/2018ApJS..237...13L} {237, 13}

\bibitem[\protect\citeauthoryear{{Marassi}, {Schneider}, {Limongi}, {Chieffi},
  {Graziani}  \& {Bianchi}}{{Marassi} et~al.}{2019}]{marassi2019}
{Marassi} S.,  {Schneider} R.,  {Limongi} M.,  {Chieffi} A.,  {Graziani} L.,
  {Bianchi} S.,  2019, \mn@doi [\mnras] {10.1093/mnras/sty3323}, \href
  {https://ui.adsabs.harvard.edu/abs/2019MNRAS.484.2587M} {484, 2587}

\bibitem[\protect\citeauthoryear{{Mart{\'\i}nez-Gonz{\'a}lez}, {W{\"u}nsch},
  {Silich}, {Tenorio-Tagle}, {Palou{\v{s}}}  \&
  {Ferrara}}{{Mart{\'\i}nez-Gonz{\'a}lez} et~al.}{2019}]{martinez2019}
{Mart{\'\i}nez-Gonz{\'a}lez} S.,  {W{\"u}nsch} R.,  {Silich} S.,
  {Tenorio-Tagle} G.,  {Palou{\v{s}}} J.,   {Ferrara} A.,  2019, \mn@doi [\apj]
  {10.3847/1538-4357/ab571b}, \href
  {https://ui.adsabs.harvard.edu/abs/2019ApJ...887..198M} {887, 198}

\bibitem[\protect\citeauthoryear{{Mathis}, {Rumpl}  \& {Nordsieck}}{{Mathis}
  et~al.}{1977}]{mathis1977}
{Mathis} J.~S.,  {Rumpl} W.,   {Nordsieck} K.~H.,  1977, \mn@doi [\apj]
  {10.1086/155591}, \href {http://adsabs.harvard.edu/abs/1977ApJ...217..425M}
  {217, 425}

\bibitem[\protect\citeauthoryear{{Matsuura} et~al.,}{{Matsuura}
  et~al.}{2015}]{matsuura2015}
{Matsuura} M.,  et~al., 2015, \mn@doi [\apj] {10.1088/0004-637X/800/1/50},
  \href {http://adsabs.harvard.edu/abs/2015ApJ...800...50M} {800, 50}

\bibitem[\protect\citeauthoryear{{Mattsson}, {Andersen}  \&
  {Munkhammar}}{{Mattsson} et~al.}{2012}]{mattsson2012}
{Mattsson} L.,  {Andersen} A.~C.,   {Munkhammar} J.~D.,  2012, \mn@doi [\mnras]
  {10.1111/j.1365-2966.2012.20575.x}, \href
  {https://ui.adsabs.harvard.edu/abs/2012MNRAS.423...26M} {423, 26}

\bibitem[\protect\citeauthoryear{{Morgan} \& {Edmunds}}{{Morgan} \&
  {Edmunds}}{2003}]{morgan2003}
{Morgan} H.~L.,  {Edmunds} M.~G.,  2003, \mn@doi [\mnras]
  {10.1046/j.1365-8711.2003.06681.x}, \href
  {http://adsabs.harvard.edu/abs/2003MNRAS.343..427M} {343, 427}

\bibitem[\protect\citeauthoryear{{Nanni}, {Bressan}, {Marigo}  \&
  {Girardi}}{{Nanni} et~al.}{2013}]{nanni2013}
{Nanni} A.,  {Bressan} A.,  {Marigo} P.,   {Girardi} L.,  2013, \mn@doi
  [\mnras] {10.1093/mnras/stt1175}, \href
  {https://ui.adsabs.harvard.edu/abs/2013MNRAS.434.2390N} {434, 2390}

\bibitem[\protect\citeauthoryear{{Nanni}, {Burgarella}, {Theul{\'e}},
  {C{\^o}t{\'e}}  \& {Hirashita}}{{Nanni} et~al.}{2020}]{nanni2020}
{Nanni} A.,  {Burgarella} D.,  {Theul{\'e}} P.,  {C{\^o}t{\'e}} B.,
  {Hirashita} H.,  2020, \mn@doi [\aap] {10.1051/0004-6361/202037833}, \href
  {https://ui.adsabs.harvard.edu/abs/2020A&A...641A.168N} {641, A168}

\bibitem[\protect\citeauthoryear{{Niculescu-Duvaz}, {Barlow}, {Bevan},
  {Milisavljevic}  \& {De Looze}}{{Niculescu-Duvaz}
  et~al.}{2021}]{niculescu2021}
{Niculescu-Duvaz} M.,  {Barlow} M.~J.,  {Bevan} A.,  {Milisavljevic} D.,   {De
  Looze} I.,  2021, \mn@doi [\mnras] {10.1093/mnras/stab932}, \href
  {https://ui.adsabs.harvard.edu/abs/2021MNRAS.504.2133N} {504, 2133}

\bibitem[\protect\citeauthoryear{{Priestley}, {Barlow}  \& {De
  Looze}}{{Priestley} et~al.}{2019}]{priestley2019}
{Priestley} F.~D.,  {Barlow} M.~J.,   {De Looze} I.,  2019, \mn@doi [\mnras]
  {10.1093/mnras/stz414}, \href
  {http://adsabs.harvard.edu/abs/2019MNRAS.485..440P} {485, 440}

\bibitem[\protect\citeauthoryear{{Priestley}, {Bevan}, {Barlow}  \& {De
  Looze}}{{Priestley} et~al.}{2020}]{priestley2020b}
{Priestley} F.~D.,  {Bevan} A.,  {Barlow} M.~J.,   {De Looze} I.,  2020,
  \mn@doi [\mnras] {10.1093/mnras/staa2121}, \href
  {https://ui.adsabs.harvard.edu/abs/2020MNRAS.497.2227P} {497, 2227}

\bibitem[\protect\citeauthoryear{{Priestley}, {Chawner}, {Matsuura}, {De
  Looze}, {Barlow}  \& {Gomez}}{{Priestley} et~al.}{2021}]{priestley2021}
{Priestley} F.~D.,  {Chawner} H.,  {Matsuura} M.,  {De Looze} I.,  {Barlow}
  M.~J.,   {Gomez} H.~L.,  2021, \mn@doi [\mnras] {10.1093/mnras/staa3445},
  \href {https://ui.adsabs.harvard.edu/abs/2021MNRAS.500.2543P} {500, 2543}

\bibitem[\protect\citeauthoryear{{Robitaille} \& {Whitney}}{{Robitaille} \&
  {Whitney}}{2010}]{robitaille2010}
{Robitaille} T.~P.,  {Whitney} B.~A.,  2010, \mn@doi [\apjl]
  {10.1088/2041-8205/710/1/L11}, \href
  {https://ui.adsabs.harvard.edu/abs/2010ApJ...710L..11R} {710, L11}

\bibitem[\protect\citeauthoryear{{Salpeter}}{{Salpeter}}{1955}]{salpeter1955}
{Salpeter} E.~E.,  1955, \mn@doi [\apj] {10.1086/145971}, \href
  {https://ui.adsabs.harvard.edu/abs/1955ApJ...121..161S} {121, 161}

\bibitem[\protect\citeauthoryear{{Sutherland} \& {Dopita}}{{Sutherland} \&
  {Dopita}}{2017}]{sutherland2017}
{Sutherland} R.~S.,  {Dopita} M.~A.,  2017, \mn@doi [\apjs]
  {10.3847/1538-4365/aa6541}, \href
  {https://ui.adsabs.harvard.edu/abs/2017ApJS..229...34S} {229, 34}

\bibitem[\protect\citeauthoryear{{Triani}, {Sinha}, {Croton}, {Pacifici}  \&
  {Dwek}}{{Triani} et~al.}{2020}]{triani2020}
{Triani} D.~P.,  {Sinha} M.,  {Croton} D.~J.,  {Pacifici} C.,   {Dwek} E.,
  2020, \mn@doi [\mnras] {10.1093/mnras/staa446}, \href
  {https://ui.adsabs.harvard.edu/abs/2020MNRAS.493.2490T} {493, 2490}

\bibitem[\protect\citeauthoryear{{Wesson}, {Barlow}, {Matsuura}  \&
  {Ercolano}}{{Wesson} et~al.}{2015}]{wesson2015}
{Wesson} R.,  {Barlow} M.~J.,  {Matsuura} M.,   {Ercolano} B.,  2015, \mn@doi
  [\mnras] {10.1093/mnras/stu2250}, \href
  {http://adsabs.harvard.edu/abs/2015MNRAS.446.2089W} {446, 2089}

\bibitem[\protect\citeauthoryear{{Woosley} \& {Weaver}}{{Woosley} \&
  {Weaver}}{1995}]{woosley1995}
{Woosley} S.~E.,  {Weaver} T.~A.,  1995, \mn@doi [\apjs] {10.1086/192237},
  \href {http://adsabs.harvard.edu/abs/1995ApJS..101..181W} {101, 181}

\bibitem[\protect\citeauthoryear{{Yamasawa}, {Habe}, {Kozasa}, {Nozawa},
  {Hirashita}, {Umeda}  \& {Nomoto}}{{Yamasawa} et~al.}{2011}]{yamasawa2011}
{Yamasawa} D.,  {Habe} A.,  {Kozasa} T.,  {Nozawa} T.,  {Hirashita} H.,
  {Umeda} H.,   {Nomoto} K.,  2011, \mn@doi [\apj]
  {10.1088/0004-637X/735/1/44}, \href
  {https://ui.adsabs.harvard.edu/abs/2011ApJ...735...44Y} {735, 44}

\makeatother
\end{thebibliography}


\appendix

\section{The metallicity scaling of dust destruction}
\label{sec:scaling}

\begin{figure}
  \centering
  \includegraphics[width=\columnwidth]{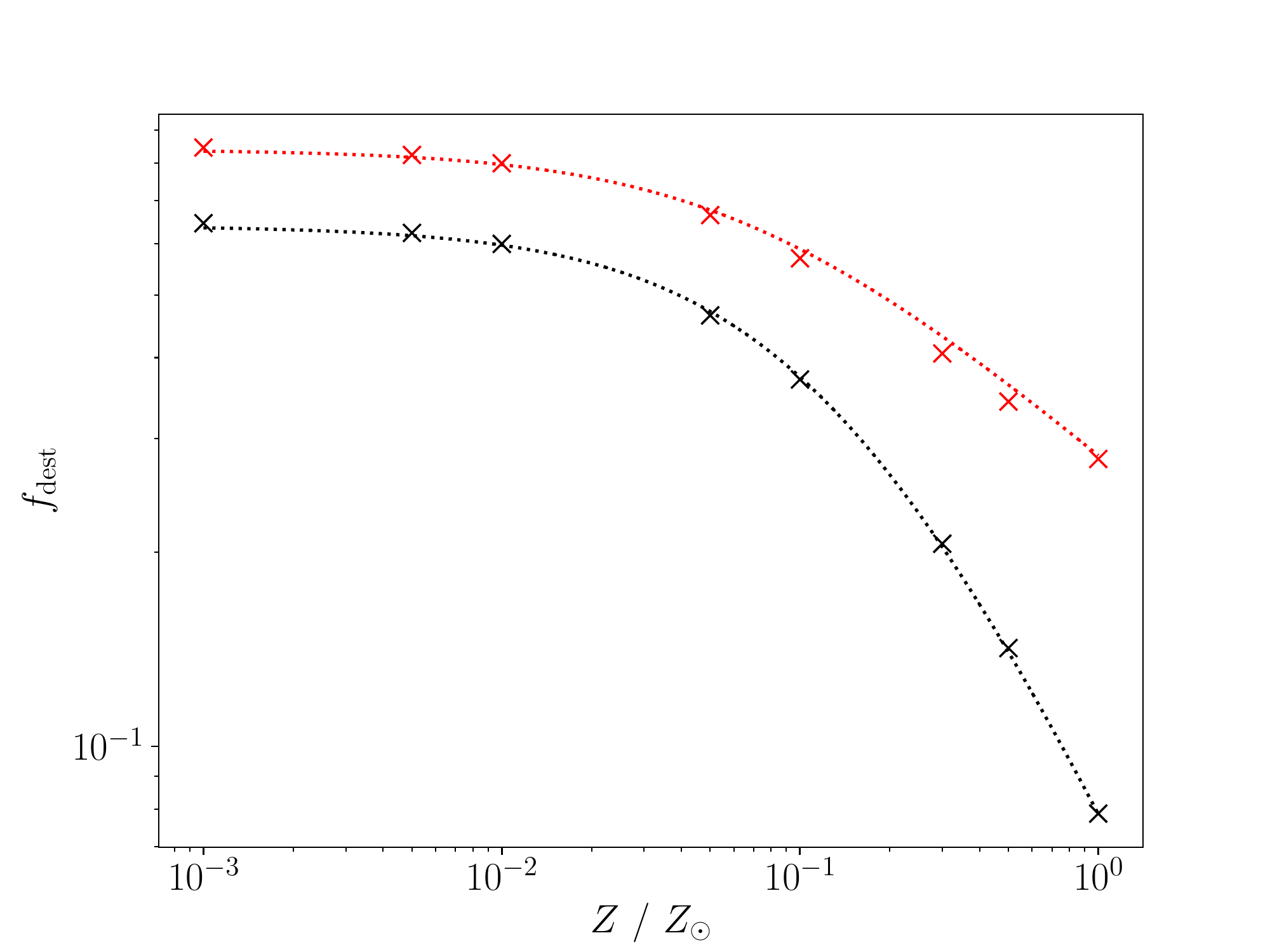}
  \caption{The dust destruction efficiency via thermal sputtering for a $200 \kms$ shock versus metallicity {(black crosses), and with an additional efficiency of $0.2$ to represent kinetic sputtering (red crosses)}. The {thermal} model efficiencies are well-reproduced {by $\fdest = 0.64 / (1 + Z / 0.14 \zsun)$ (black dotted line), the increased values by $\fdest = 0.84 / (1 + Z / 0.07 \zsun)^{0.4}$ (red dotted line).}}
  \label{fig:fdest}
\end{figure}

{We calculate the evolution of plane-parallel shocks in a $1 \pcc$ ambient medium of varying metallicity using {\sc mappings} \citep{sutherland2017}. We use the gas density and temperature (as a function of time post-shock) to calculate the thermal sputtering rate, $\frac{da}{dt}$, of silicate grains following the prescription in \citet{biscaro2016}. We then integrate $\frac{da}{dt}$ from $t=0$, when the shock first impacts the ambient medium, to the time where the gas has cooled below $10^4 \kel$. This gives the reduction in grain size, $a_{\rm sput}$, due to thermal sputtering in the shocked gas. We determine the destruction efficiency, $\fdest$, by taking a \citet{mathis1977} $\frac{dn}{da} \propto a^{-3.5}$ grain size distribution and calculating the reduction in dust mass if every grain has its size reduced by $a_{\rm sput}$.}

Figure \ref{fig:fdest} shows the variation of $\fdest$ with metallicity for a $200 \kms$ shock impacting a $1 \pcc$ ambient medium. {The results are well-fit by $\fdest = 0.64 / (1 + Z / 0.14 \zsun)$. We interpret the $\fdest \sim Z^{-1}$ behaviour at high $Z$ as due to the approximately linear relationship between the cooling rate and the metallicity for temperatures $\gtrsim 10^5 \kel$, where cooling is dominated by metal line emission. The cooling timescale is then $t_{\rm cool} \sim Z^{-1}$, so the period of time where thermal sputtering is effective also varies as $Z^{-1}$, and if $\frac{da}{dt} \sim {\rm const.}$ then $a_{\rm sput} \sim Z^{-1}$ (we have confirmed that this is in fact the case in our implementation). Assuming a single grain size $a_0$ for simplicity, $\fdest = 1 - (a_0 - a_{\rm sput})^3/a_0^3$ which, for $a_{\rm sput} \ll a_0$, reduces to $\fdest \sim 3 a_{\rm sput}/a_0$. As the majority of the mass in a \citet{mathis1977} distribution is in the largest grains, where $a_{\rm sput} \ll a_0$ holds (at least for a $200 \kms$ shock), $\fdest \sim Z^{-1}$ until the point where metal line cooling becomes unimportant, beyond which $\fdest \sim {\rm const.}$.}

{Obtaining the mass of dust destroyed per SN from the dust destruction efficiency of single-shock models requires assumptions about the typical evolution of supernova remnants and their surrounding ISM, which may not accurately describe the complex behaviour of real objects \citep{chawner2020b,priestley2021}. Rather than attempt to derive a value of $\mclr$ from $\fdest$, we treat the normalisation of $\mclr$ (the value at some reference metallicity) as a free parameter, and assume that it follows the same scaling with $Z$ as $\fdest$ {(i.e. $\mclr \propto (1 + Z / 0.14 \zsun)^{-1}$)}. Because the physical process causing this scaling - the increase in the cooling timescale with decreasing $Z$ - is only weakly affected by the shock velocity and ambient density, we expect that it should be preserved regardless of how a value of $\mclr$ is obtained from $\fdest$.}

{Our model neglects kinetic sputtering, which does not have any obvious dependence on metallicity. If we add a constant value of $0.2$ to $\fdest$ at all metallicities to approximate this effect, shown in Figure \ref{fig:fdest}, we find a much shallower $\fdest \sim Z^{-0.4}$, close to the $-0.298$ exponent from Y11. The $\fdest \sim Z^{-1}$ relation derived from thermal sputtering may thus represent an extreme case, which we use in this paper to assess the potential impact, although we argue in Section \ref{sec:discussion} that grain shattering (not included by Y11) might result in a stronger metallicity dependence than sputtering alone.}


\bsp	
\label{lastpage}
\end{document}